\documentclass{mn2e}
\usepackage{times}

\input{psfig.sty}

\newif\ifAMStwofonts

\title[Coronal outflow dominated accretion discs]
        {Coronal outflow dominated accretion discs: a new possibility for low luminosity black holes?}
\author[A. Merloni \& A. C. Fabian]
        { A. Merloni and  A. C. Fabian
\\Institute of Astronomy, Madingley Road, Cambridge, CB3 0HA
}

\date{}

\begin{document}

\maketitle

\label{firstpage}

\begin{abstract}
The spectral energy distributions of galactic black holes in the low/hard state and of low-luminosity AGN
possess many common features, the most prominent being: compact, flat (or inverted) spectrum radio cores
with high brightness temperatures; excess red and infrared emission, often correlated with the radio flux;
an extremely weak (or absent) quasi-thermal hump and a hard X-ray power-law with high energy cut-off.
These sources are thought to be accreting at low rates and advection (or convection) dominated accretion flows
are usually considered the best candidates to explain them. Here we present an alternative
possibility, involving strong, unbound, magnetic coronae generated by
geometrically thin, optically thick accretion discs at low accretion rates.
First we show that, if angular momentum transport in the disc is due to magnetic
turbulent stresses, the magnetic energy density and effective viscous stresses inside the disc
are proportional to the geometric mean of the total (gas plus radiation) and gas pressure. Therefore the corona
is less powerful in a radiation pressure dominated disc, and the relative fraction of the power liberated in
the corona increases as the accretion rate decreases. Furthermore, we discuss reasons why
energetically dominant coronae are ideal sites for
launching powerful jets/outflows, both MHD and thermally driven.
In analysing the spectral properties of such coronal outflow dominated accretion discs, we reach
the important conclusion that if the jet/outflow is, as is likely, radiatively
inefficient, then so is the source overall, even without advection of energy into the black hole
being relevant for the dynamics of the accretion flow.
\end{abstract}


\section{Introduction: ADAF or Jets ?}
The spectral energy distribution (SED)
of the gravitational power released as electromagnetic radiation when matter accretes onto
a compact object is far from universal. Different accretion modes are possible, and often the
same initial conditions at the outer boundaries admit more than one solution for
the accretion flow configuration at the inner boundary, with often very different
radiative properties. The main goal of accretion flows theory is to
distinguish and understand all the possible
different modes of accretion, and classify the observed sources in terms of
 such modes.

In the case of non-quiescent
Galactic (stellar mass) Black Holes Candidates (GBHC), an X-ray based
classification in terms of their {\it spectral states} has emerged in recent years.
Broadly speaking, the different states are distinguished by their broadband luminosity,
and by the relative contribution to the luminosity
of a `soft', quasi-thermal and a `hard', power-law like spectral components (see e.g. the recent
review by Done 2001 and reference therein).
At luminosities close to the Eddington one, these sources are often in the {\it Very High State},
where both of the two components contribute substantially to the SED. At slightly lower
luminosities the quasi-thermal one dominates, and the power-law is usually steeper and
extended  to the gamma-ray band ({\it High/Soft State}). At even lower luminosities
the spectra are completely dominated by a hard power-law component, with the quasi-thermal
one extremely weak or even absent: these are the so-called {\it Low/Hard States}.
Sometimes, at luminosities intermediate between those of the Soft and the Hard States, an
{\it Intermediate State} is observed, with properties similar to those of the Very High State.
It is often the case that the same source, either persistent or transient, undergoes
a transition between spectral states, and therefore accretion modes.

From the theoretical point of view, the quasi-thermal component is usually associated with
a geometrically thin, optically thick accretion disc. Its dynamical and geometrical properties
could in principle be deduced solely by a careful comparison between the observed spectral
properties of the quasi-thermal hump and a detailed model for the accretion disc.
On the other hand, the observed hard X-ray power-laws represent a universal
signature of a specific
physical process (most likely inverse Compton scattering of soft photons on a
population of hot electrons; Sunyaev \& Titarchuk 1980) more than of specific accretion dynamics.
 This is the reason why, if there is little doubt that the standard accretion
disc model (Shakura \& Sunyaev 1973) captures the basic physical properties of black holes in their
High/Soft state, the accretion mode responsible for the Low/Hard state is still matter of debate.

Recently, multi-wavelength observations of GBHC in the Low/Hard state have been adding important
pieces of information, which are helping to shed more light on the subject. Crucial in this
respect is the discovery of the high-galactic latitude transient source XTE J1118+480.
This source was persistently in the Low/Hard state during its 2000 outburst, and being
very little affected by galactic extinction, has provided the most accurate and complete
SED to date of a galactic black hole in this  state \cite{mcc01,fro01}.
It possesses all the typical characteristics that distinguish such state \cite{zdz00,fen01},
namely:
a hard X-ray power-law with photon index $\Gamma\simeq 1.8$ and a high energy cut-off,
indicative of thermal Comptonization; a dim and cold thermal disc component peaking in the
EUV band \cite{mcc01};
 excess red and infrared emission compared the one expected from standard disc models;
a compact (unresolved) flat (or inverted) radio core, probably extending to the NIR or optical
regime \cite{fen01b}.

The situation for supermassive black holes accreting in the centre of active galactic
nuclei bears some resemblance with the GBHC one.
Beside the best studied case of the black hole at the centre of our Galaxy (Melia \& Falcke 2001, and
references therein),
recent multi-wavelength campaigns  on nearby low-luminosity AGNs (LLAGN)
(Ho 1999, Nagar et al. 2000, Falcke et al. 2000, Ho et al. 2001) have
revealed nuclear SEDs markedly different from the canonical broadband continuum spectrum of
high luminosity AGN (as compiled, e. g.,  by Elvis et al. 1994). The most striking
difference is the lack of UV excess, the so-called `big blue bump' associated with
the emission from an optically thick, geometrically thin accretion disc \cite{ho99}. In addition,
VLA \cite{nag00} and VLBA \cite{fal00} radio observations of LLAGN
have revealed flat or inverted compact radio cores, with an incidence higher than in the case of
normal Seyfert galaxies, and, for the sources observed at higher spatial resolution with VLBA,
very high brightness temperatures ($T_b \ga 10^8$K).

In light of the above observational similarities between the two classes,
we would like to discuss here a possible answer to the following question:
what is the common accretion mode for low-luminosity black holes?

Since their rediscovery in recent years (Narayan \& Yi 1994), radiatively inefficient, optically
thin, advection dominated accretion
flows (ADAF; Ichimaru 1977, Rees et al. 1982, Abramowicz et al. 1995;
see Narayan, Mahadevan \& Quataert 1999 for a recent review)
have been regarded as natural solutions.
They can explain the absence of the quasi-thermal hump,
the low X-ray luminosities and the observed X-ray spectral slopes.
When tested against the best data for GBHC in their Low/Hard state, though,
as in the case of XTE J1118+480 \cite{es01} or
Cygnus X-1 \cite{es98},
ADAF models alone cannot work. A transition between an inner ADAF and an outer
Shakura Sunyaev disc is needed, with $R_{\rm tr} \approx 10 - 100 R_{\rm S}$, as can also be
inferred from studies of X-ray reflection components
(Gilfanov, Churazov \& Revnivtsev 2000; Done 2001, and reference therein).
Detailed spectral fits to the broadband SED of the nearby LLAGNs M81 and NGC 4579
(Quataert et al. 1999) lead to similar conclusions. The physics
of such a transition, which may be taking place through a gradual evaporation
of the disc as proposed by Meyer \& Meyer-Hofmeister (1994) or via turbulent diffusive heat transport
in the radial direction \cite{hon96},
is not included in the ADAF models, so that $R_{\rm tr}$ is usually treated as a free
parameter in the spectral fit procedure.
Furthermore, the low energy part of the SED is also difficult to
reconcile with the standard ADAF picture, that in general predicts highly inverted
spectra, instead of the observed flat ones.
In fact, morphology and spectral indices of the compact radio cores require additional outflows
from the central inner part in order to be consistent with the prediction of
an adiabatic flow \cite{dcf01}. Finally, we must note that, both from the theoretical
point of view \cite{ali00} and from
numerical simulation \cite{ian00}, it is clear that radiatively inefficient flows
are subject to strong convective instabilities
(and might therefore be better called CDAF;
Stone, Pringle \& Begelman 1999, Quataert \& Gruzinov 2000a), but are not in general able to generate
strong outflows (although, see Misra \& Taam 2001), unless the viscosity is very high.

Flat or slightly inverted radio spectra from very compact
sources are generally associated with jet models \cite{bk79,rey82}.
This has prompted the alternative view  that spectra of LLAGN are essentially
jet-dominated \cite{fal01}. As a matter of fact, in low-luminosity
nearby radio galaxies observed with HST \cite{cap00}
compact optical cores seem very common. Chiaberge et al. (1999), analysing
HST images of a complete sample of 33 FR I radio galaxies, demonstrated that the luminosity
of compact optical cores in these sources correlates linearly with that of the radio cores,
suggesting a common non-thermal synchrotron emission origin for both.
In particular, detailed spectral jet models
have been demonstrated to successfully reproduce the SED of Sgr A$^*$ (Falcke \& Markoff 2000).
Interestingly, this might also be the case for the X-ray transient XTE J1118+480,
\cite{mff01}. The presence of a relativistic jet in the black hole candidate Cygnus X-1
in the low/hard state has been recently confirmed by milliarcsecond resolution VLBA observations
\cite{sti01}, while Fender (2001),
on the basis of the observational evidence of strong radio compact jet
correlating with hard X-ray emission in GBHC in the Low/Hard state, concluded that,
in these sources, the jet power is at least comparable with the accretion one.

Here we explore further this latter hypothesis from the point of view of the standard accretion
disc theory. We shall propose a possible new model for low-luminosity black holes, in
which a standard geometrically thin, optically thick disc at low accretion rates
dissipates a large fraction of its gravitational energy in a magnetic corona.
We shall outline the reasons why energetically dominant coronae are the ideal site for
launching powerful jets/outflows, and discuss the expected spectral properties of such
accretion flows. The important conclusion we reach is that if the jet/outflow is radiatively
inefficient, then so is the source overall.

\section{Coronal strength at low accretion rates}
\label{sec_f_mdot}

It is well known that micro-physical viscosity is in general too inefficient to drive
accretion at rates comparable to the observed ones, and enhanced/turbulent viscosity is
therefore a theoretical necessity. As astrophysical discs are bound to be seeded with
magnetic fields, magneto-rotational instability (MRI) \cite{vel59,cha60,bh91} is the best
candidate to date to produce the self-sustained turbulence that is able to transport
angular momentum in rotationally supported discs. Any weak, sub-thermal magnetic field,
coupled with outwardly decreasing differential rotation (two conditions easily met by
any accretion flow endowed with angular momentum) generate the turbulence which in turn
is able to regenerate the field \cite{tp92}. 
This instability, which has a rapid growth rate of the
order of the orbital frequency $\Omega$, results in a greatly enhanced effective viscosity
that is able to transport angular momentum outward (Balbus \& Hawley 1998).

What is the fate of the magnetic field which is amplified inside the disc?
Two competing mechanisms should operate in an accretion  disc to saturate the field:
dissipation and buoyancy.
On one hand, part of the field {\it must} be dissipated,
if the turbulence has to be sustained and MRI relied upon to
transport angular momentum. In fact it has been shown \cite{bh98} that,
in a steady state disc, if angular momentum
transport is due solely to  magneto-rotational instability, magnetic fields must
be sub-thermal. This in turn implies that buoyancy cannot be a too efficient
mechanism to get rid of the disc field:
to cross one disc scaleheight in less than a dynamical time $1/\Omega$
a velocity of the order of the sound speed is required, which would
correspond to equipartition magnetic field. Such strong fields would suppress
the MRI \cite{bh98}.

On the other hand, we  can expect  the magnetic
field configuration in the disc to be highly intermittent \cite{pp95},
with the field
concentrated in small regions (flux tubes and ropes)
with relatively low filling factor (see e.g. Blackman 1996 and the
simulations presented in
Machida, Hayashi \& Matsumoto 2000). These structures
can more easily rise buoyantly
and emerge in the low density part of the flow
above the accretion disc to form a magnetic atmosphere, the corona,
where the field is finally dissipated.

It is not clear at present what is the relative importance of the two mechanisms
of field saturation inside a standard accretion disc, if
they either coexist or instead alternate in a cyclic fashion
(as, for example, in the sequence: low disc field $\rightarrow$ MRI $\rightarrow$ stronger field $\rightarrow$
buoyancy $\rightarrow$ field expulsion $\rightarrow$ suppression of the MRI
$\rightarrow$ low disc field $\rightarrow$ MRI; see also the discussion in 
Tout \& Pringle, 1992).

Numerical simulations of vertically stratified discs \cite{bnst,shgb} seem to suggest
that, although dissipation is the main saturation mechanism for the instability,
buoyancy is indeed capable of transporting vertically a substantial
fraction of magnetic energy to power a strong corona (see in particular Miller \&
Stone 2000, where a larger vertical extent is explored), without
quenching the magneto-rotational instability in the underlying disc.

In this section we investigate how changes in the internal
disc structure, associated with variation in the external accretion rate,
may influence the accretion disc--corona system.
To this end we only consider standard
optically thick and geometrically thin accretion discs
sandwiched by a patchy magnetic corona \cite{grv79,hmg94,dcf99,mf01b}.
 We will assume that buoyant vertical transport of magnetic flux tubes is a relatively efficient
way of disposing of the disc field, and discuss the self-consistency
of such assumption at the end of the section.

The nature of the saturation mechanism determines the level up to which magnetic fields are amplified 
inside the disc. Many authors have argued that
magnetic field amplification is likely to be limited to values such that
the magnetic pressure is at most equal to the local {\it gas} pressure, even in its
inner, radiation pressure-dominated parts \cite{sc81,sr84,sc89,vis95a}.  
In a recent paper Blaes \& Socrates (2001) have analysed local
dynamical instabilities in magnetized, radiation pressure-supported (and geometrically
thin) accretion discs, and found that, due to radiative diffusion,
 the growth rate of the MRI is reduced in such
systems to $ \sigma \sim \Omega (c_{\rm g}/v_{\rm A})$, where $c_{\rm g}$ is the gas sound speed
and $v_{\rm A}$ is the Alfv\'en speed, well below the usual rapid growth rate ($\sim
\Omega$) in gas pressure dominated discs. 
We can therefore give an order of magnitude estimate of the amplitude of the field
in the radiation pressure dominated part of a standard accretion disc. If we assume that the buoyant 
vertical speed of a magnetic flux tube is approximately given by the Alfv\'en speed, we
get for the buoyant timescale $t_{\rm b}\sim H/v_{\rm A} \simeq c_{\rm s}/(\Omega v_{\rm A})$,
 $c_{\rm s}=\sqrt{P_{\rm tot}/\rho}$ is the isothermal sound speed. The magnetic field,
with initial (subthermal) amplitude $B_0$, rapidly grows to $B \sim B_0 e^{\sigma t_{\rm b}}$. 
Therefore we have $\ln(B/B_0) \simeq c_{\rm s}c_{\rm g}/v_{\rm A}^2$. Neglecting the logarithmic
dependence on the initial field, this in turn suggests that the field will saturate at 
an amplitude such that
\begin{equation}
\frac{B^2}{8\pi}=P_{\rm mag} \simeq \alpha_0 \sqrt{P_{\rm gas}P_{\rm tot}},
\end{equation}
where $P_{\rm tot}$  and $P_{\rm gas}$ are the total (gas plus radiation) and the gas pressure 
at the disc midplane, respectively, and $\alpha_0$ is a constant of the order of unity. 
Such a scaling was first suggested by Taam \& Lin (1984), and its 
stability properties have been studied by Lin \& Shields (1986) and, 
for the case of slim discs at high accretion rates, by Szuszkiewicz (1990).

If magnetic turbulence is ultimately responsible for transporting angular
momentum, we can still retain the usual modified $\alpha$ viscosity prescription
in which the stress tensor is proportional to the gas
pressure $\alpha = P_{\rm mag}/P_{\rm gas}$ \cite{sc81,sr84}, 
but now with $\alpha \simeq \alpha_0 \sqrt{P_{tot}/P_{\rm gas}}$. 

 As in many other previous treatments of the two-phase thermal models for
accretion flows onto compact objects
(Haardt \& Maraschi 1991; Svensson \& Zdziarski 1994), we parameterize the
coronal dominance by means of the fraction $f$ of gravitational power
associated with the angular momentum transport
that is stored in the form of magnetic structures (loops, flux tubes),
transported vertically through the disc in a dissipationless fashion,
and finally
dissipated in the corona. Thus, if the total released power
of the accretion disc--corona system is given by $L\equiv \dot m L_{\rm Edd} =
4 \pi G M m_{\rm p} \dot m c/\sigma_{\rm T}$, we define the corona luminosity as
$L_{\rm c}=f \dot m L_{\rm Edd}$.
The fraction $f$ of power dissipated in the corona can be expressed as
the ratio
of the magnetic Poynting flux in the vertical direction to the
locally dissipated energy flux in the disc:
\begin{equation}
\label{f1}
f = \frac{F_{\rm P}}{Q}=\frac{P_{\rm mag} v_{\rm P}}{Q},
\end{equation}
where $F_{\rm P}$ is the vertical Poynting flux and $Q$ is the dissipation
rate of gravitational energy per unit surface area, and we have defined the
velocity $v_{\rm P}$ with which the magnetic flux is transported
in the vertical direction.
From the standard equations of geometrically thin accretion discs
with modified viscosity law we have
\begin{equation}
Q=\frac{3}{2}\alpha v_{\rm K} \left(\frac{H}{R}\right) P_{\rm gas} =
\frac{3}{2}\alpha c_{\rm s} P_{\rm gas},
\end{equation}
where $v_{\rm K}$ is the Keplerian velocity.

The main source of uncertainties lies in the complex process of vertical transport of
magnetic flux tubes in the body of the accretion disc, and therefore on the value of the
velocity $v_{\rm P}$ (see Coroniti 1981; Stella \& Rosner 1984; Vishniac 1995a,b;  for a
more thorough discussion). Here we assume that the rise speed of the magnetic tubes
is proportional to their internal  Alfv\'en speed, and consequently scales as $v_{\rm
P}/c_{\rm s}= b \beta^{-1/2}$, where $\beta=P_{\rm tot}/P_{\rm mag}$
 is the standard plasma parameter, and $b$ is related to the efficiency of
buoyant transport of magnetic structure in the vertical direction inside the disc, which
is of the order of unity for extremely evacuated tubes.

Equation~(\ref{f1}) then becomes
\begin{equation}
\label{eq_f2}
f \simeq \frac{2 b}{3} \beta^{-1/2} = \frac{2 b \sqrt{\alpha_0}}{3} \left(1+\frac{P_{\rm rad}}{P_{\rm gas}}\right)^{-1/4}.
\end{equation}

The fraction $f$ can then be calculated for any given accretion rate and at any given
radius. If the mass of the central black hole is $M=m M_{\odot}$ and the radial distance
from the central source is expressed in units of Schwarzschild radii $r=R/R_{\rm s}= R
c^2/2GM$, from (\ref{eq_f2}) and the equations for the ratio $P_{\rm
rad}/P_{\rm gas}$ with modified viscosity law, we obtain the following implicit equation
for $f(r)$ as a function of $r$ in the radiation and gas pressure dominated zones of the disc, respectively
(cfr. Stella \& Rosner 1984, appendix A)\footnote{We have derived the structure
equations for a cold disc on the basis of Svensson \& Zdziarski 1994, eqs. 1-4. The
parameter $\xi$ of the radiative diffusion equation has been set equal to 1 for the
radiation pressure dominated solution and to 8/3 in the gas pressure one in order for
the two solutions to match continuously at the boundary. Also, for the ease of computation, we have approximated
$(\alpha/\alpha_0)^{1/5} \simeq 1$ and $(\alpha/\alpha_0)^{1/10} \simeq1$ in the expression for $P_{\rm rad}/P_{\rm gas}$.}
\begin{equation}
\label{eq_f_rad}
\frac{Kf(r)^{-4}-1}{(1-f(r))^{9/5}} \simeq 2.01 \times 10^5 (\alpha_0 m)^{1/5}
r^{-21/10} [\dot m J(r)]^{8/5}
\end{equation}
\begin{equation}
\label{eq_f_gas}
\frac{Kf(r)^{-4}-1}{(1-f(r))^{9/10}} \simeq 1.31 \times 10^2 (\alpha_0 m)^{1/10}
r^{-21/20} [\dot m J(r)]^{4/5},
\end{equation}
where the function $J(r)=1-\sqrt{3/r}$ comes from the no-torque inner boundary condition.
The numerical factor $K=(2b\sqrt{\alpha_0}/3)^4$ contains the unknown quantities
in our treatment. As we shall see, the value of $K$ is related to the maximal
fraction of power released in the corona. It is evident that 
high viscosities and vertical rise speeds (that are associated with the presence of very low-$\beta$,
strongly evacuated, magnetic filaments in the disc)
are needed in order for the corona to be powerful.

The fraction of power dissipated in the corona is higher when gas pressure
dominates over radiation pressure inside the disc.
This happens over a larger and
larger portion of the disc as the accretion rate decreases.
The radial boundary between the two solution is found when the fraction
of power released in the corona is $f(r)=(K/2)^{1/4}$, 
and can be calculated by solving the equation
\begin{equation}
\frac{r}{J(r)^{16/21}} \simeq 350 (\alpha_0 m)^{2/21} \dot m^{16/21} (1-(K/2)^{1/4})^{6/7}.
\end{equation}
The left hand side has a minimum at $r_*=5.72$, therefore, for accretion rates smaller than
a critical value, $\dot m_{\rm cr} \simeq 0.016 (\alpha m)^{-1/8} (1-(K/2)^{1/4})^{-9/8}$,
the disc is gas pressure dominated even in its inner part.

\begin{figure}
\psfig{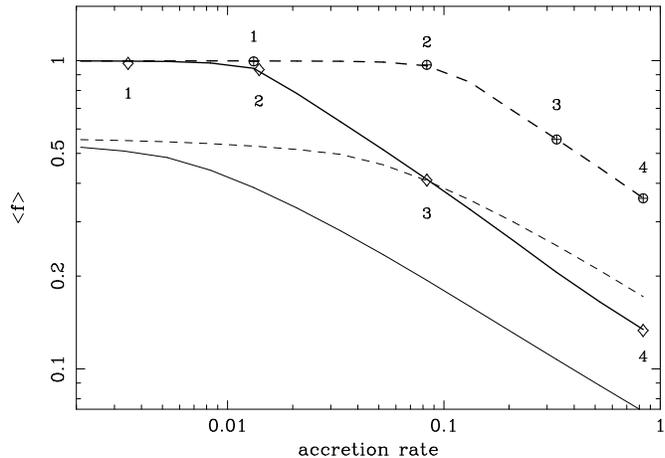}
\caption{The integral fraction of accretion power dissipated into the corona
as a function of the accretion rate for $\alpha_0=0.3$ and two different values of
central black hole
mass: $M=10^8 M_{\odot}$ (solid lines) and $M=10 M_{\odot}$ (dashed lines), and
of the coefficient
$K$: thicker lines corresponds to $K=0.99$ and thinner ones to $K=0.1$. The maximal
coronal strength is reached for the smallest accretion rates, and is roughly
equal to $K^{1/2}$. The marked numbers on the $K=0.99$ curves
correspond to the spectral energy distributions shown
in Figures~\ref{spectra_10} and \ref{spectra_108}.}
\label{f_mdot_coeff}
\end{figure}

Equations (\ref{eq_f_rad}) and (\ref{eq_f_gas}) can be solved numerically to calculate
the disc structure at every radius and for every accretion rate. A global value of
$\langle f \rangle$ can then be obtained by integrating over all the disc area: 
\begin{equation}
\langle f \rangle = \frac{\int_3^{\infty} f(r) Q(r) 2 \pi r dr}
{\int_3^{\infty} Q(r) 2 \pi r dr}.
\end{equation}
In Figure~\ref{f_mdot_coeff} we plot such global value of $\langle f \rangle$ 
as a function of the total
accretion rate for two different values of $K$ (0.99 and 0.1) for typical stellar mass
and supermassive black holes.

As already pointed out, as the radiation pressure dominated part of the disc shrinks, the
value of $\langle f \rangle$ increases as a larger portion of the disc is gas pressure 
dominated. For accretion rates smaller than the critical one, $\dot m_{\rm cr}$, when 
the disc is completely gas pressure dominated, the $\langle f \rangle$--$\dot m$ relation 
flattens, and the fraction of the power released in the corona increases only slightly 
as the disc becomes denser and the radiation pressure further decreases.
The maximum coronal dominance is attained for the lowest values of the
accretion rate, and is roughly equal to $K^{1/4}$. If $K>1$
eqs. (\ref{eq_f_rad}) and (\ref{eq_f_gas}) do not admit a solution with
$f(r)<1$ over a wide range of radii. This might correspond to a situation in
which the magnetic field is so strong and/or the vertical transport of magnetic
flux tubes is so efficient that the standard
accretion disc -- corona configuration in no longer stable and the
accretion disc, which cannot maintain its Keplerian profile,
is effectively fragmented by the magnetic field into a number
of disconnected rings (Heyvaerts \& Priest 1989). Also, very low-$\beta$ discs are
 qualitatively different in terms of angular momentum
transport processes. In fact, in such cases the magneto-rotational instability
is suppressed and magnetic stresses at the disc--corona
interface may be the primary angular momentum vehicle (Blandford \& Payne 1982)
and the standard disc structure we have assumed may not be appropriate. We therefore did not
investigate such a case further, restricting ourselves to the stable case in which
$\beta \ga 1$ in the disc, and $K\la 1$.

\begin{figure}
\psfig{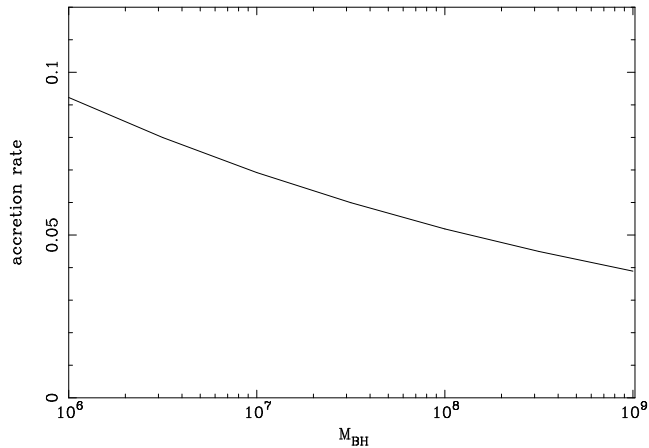}
\caption{As a function of the central black hole mass, we plot the value of 
the accretion rate for which the dissipation in the corona is more than the internal 
disc one ($f>0.5$).
}
\label{mass}
\end{figure}

The coronal strength depends weakly on the central source mass,
as implicitly shown in eq.~(\ref{eq_f_rad}) and (\ref{eq_f_gas}).
In Fig.~\ref{mass} we show explicitly the values of the accretion rate
for which $f>0.5$ ($\dot m_{\rm 5,f}\propto M_{\rm BH}^{-0.124}$).
Ghisellini \& Celotti (2001b), by studying the dividing line between FR I and FR II radio-galaxies (which
correspond approximately to low and high radio--power galaxies, respectively) in the
radio--host galaxy luminosity plane conclude that such dividing line correspond to an approximately
universal (independent on the black hole mass) transition value of $\dot m \simeq 0.06$,
consistent with the values we found for the accretion rates at which the corona starts to be
the dominant repository of gravitational energy.

To summarize, in accretion discs where angular momentum is transported
either by magnetic turbulence, or stresses,
magnetic coronae should be dominant at low accretion rates, and
their strength depends upon the nature of magnetic dissipation inside the disc. Furthermore,
magnetic coronae are stronger in stellar mass black holes than in AGN
(see also Fig.~\ref{mass} below).

\section{The fate of the coronal energy:
coronal heating and Jet launching mechanisms}

What is the fate of the energy deposited in the corona by the magnetic
structures that emerge from the disc when $f$ is large?\footnote{For the sake of clarity, in the rest of this work we will use $f$ to indicate the global value of the fraction of gravitational power released in the corona $\langle f \rangle$, neglecting any radial dependence.} 
If the current paradigm for high energy emission from black holes is correct, involving
inverse Compton scattering of soft photons on a population of thermal electrons at
temperatures $T_{\rm e}\sim 10^9$K,
then part of this energy must be used to heat the
corona, most likely via magnetic reconnection \cite{dm98} and then efficiently radiated away.
On the other hand, because Compton cooling is so efficient in the strong radiation field
of an accretion disc, the energy densities of the thermal electrons and of the
hard X-ray radiation are likely to be only a small fraction of the total
energy stored in the corona, whose dominant contribution should come from
stored magnetic field (Merloni \& Fabian 2001a) and possibly hot ions, if the corona is
two-temperature (Di Matteo, Blackman \& Fabian 1997). In such a situation,
the condition can be favourable for the occurrence of strong coronal outflows\footnote{As we do not discuss either the collimation
mechanism or the final velocities of the outflowing gas, but only the power channeled in
such a jet/outflow, we do not make any distinction between
jets and outflows, and use the two terms
indifferently.}.
The nature of the outflow depends crucially on the magnetic field strength and
topology and on the physical properties of the particle heating mechanisms.
 
In Merloni \& Fabian (2001a) it has been shown how the values of coronal temperatures and optical depths 
inferred from the observations of low luminosity black holes require that the main
reservoirs of energy in the corona are either a structured magnetic 
field or the protons in a two-temperature plasma. The extent to which the protons may act   
as repository of most of the dissipated energy crucially depends on how efficient  
the heating of electrons in the corona is. 
Great theoretical effort has been devoted recently to the study of particle energization
in accretion flows, mainly for its relevance for the ADAF scenario.
Quataert (1998) and Quataert \& Gruzinov (1999) have shown that in optically thin, 
advection dominated accretion flows, as long as the MHD turbulence is dissipated on 
small scales via incompressible modes (Alfv\'en waves), protons will be mainly heated 
only in high $\beta$ plasmas ($\beta \ga 10$), while electrons are more likely to be 
heated in low $\beta$ plasmas. On the other hand Blackman (1999)  argued that if modes
of dissipation involving compression dominate, then, even for $\beta \sim O(1)$, 
electron heating is negligible. Furthermore, we stress that the field configuration in the
corona ought to be more complicated than the uniform field considered by Quataert (1998) and
Quataert \& Gruzinov (1999), with the field concentrated in loops and tubes, 
similarly to those
 observed in the sun. Magnetic reconnection may play an important role in this case
(see e.g. Bisnovatyi-Kogan \& Lovelace 1997; Di Matteo 1998). Unfortunately, 
the physics of particle energization in magnetic reconnection is still poorly understood, and no clear conclusions can be safely drawn at present.
However, the conclusion that a powerful corona is the ideal launching platform for
powerful jets remains unscathed: 
as a matter of fact, magnetically dominated coronae
($\beta \la 1$) may power MHD driven outflows which can carry away a substantial fraction of
the gravitational power (see section \ref{sec_mhd}), even if the ions are subvirial. 
On the other hand,  
for $\beta \ga 1$, ions could be the main repository of the power dissipated in the corona; 
they can be heated to supervirial temperature and launch powerful thermally-driven 
outflows (see section \ref{sec_th}). 

For the sake of simplicity, we keep the treatment of these two cases separate, 
bearing in mind that in a realistic situation they may both contribute to the determination
 of the total outflowing power. We demonstrate that the power carried away
by the outflow/jet may be a dominant fraction of the total accretion power in both cases. 

To be fairly general, we consider that
the total coronal power generated by the accretion disc, $L_{\rm c}=f\dot m L_{\rm Edd}$,
can be either dissipated locally to heat the corona, and ultimately radiated away
as hard X-ray radiation with a luminosity $L_{\rm H}=(1-\eta)L_{\rm c}$, or
 used to launch a jet with power $L_{\rm j}=\eta L_{\rm c}$. In the following, we 
investigate the dependence of the parameter $\eta$ on the disc parameters.

The coronal magnetic field intensity depends on the dissipation rate and on the
field geometry. Here we assume, on the basis of the strong spectral and variability
arguments \cite{hmg94,dcf99,mf01a},
that  the corona is structured and the magnetic field concentrated in a number of
small active regions (where dissipation occurs).
The size, height and distribution of such regions determines
the details of the spectral and temporal properties of the high energy emission in
these sources (Merloni \& Fabian 2001b).
Here it will suffice to notice that a relatively small
number ($N\sim 100$, see Merloni \& Fabian 2001b)
of spherical active regions of size and height of the order of a few Schwarzschild
radii ($R_a \sim H_a \sim 1 R_{\rm S}$)
is what is needed to reproduce the average properties of the hard X-rays emission of
GBHC in their hard state and of Seyfert 1s and low luminosity AGN,
namely a spectral index $\Gamma \sim 2$, high energy cut-off around 100 keV
 and strong variability.

We therefore assume that the magnetic energy density in the corona is
\begin{equation}
\label{eq_b_lh}
\frac{B^2}{8\pi}=\frac{3 L_{\rm H}}{4\pi R_a^2 N c}\left(\frac{c}{v_{\rm diss}}\right),
\end{equation}
where $v_{\rm diss}$ is the dissipation velocity and depends on the uncertain nature of the
reconnection process \cite{dm98,mf01b}, and can be assumed to be of the order
$v_{\rm diss} \sim 0.01 c$.

\subsection{The jet power I. MHD launching mechanisms}
\label{sec_mhd}

Here we discuss the most general mechanisms by which a coronal magnetic field
can power strong outflows (either in the form of Poynting flux or in the form of a
magnetically driven material wind).

Models and simulations of jet production \cite{bz77,bp82,me99} show that it
is the {\it poloidal} component of the magnetic field which mainly drives the
production of powerful jets, and the output power\footnote{Based on the conclusion
of the analysis presented in Livio, Ogilvie \& Pringle (1999),
we will neglect the contribution to the total
power output from the energy extracted from the black hole via the Blandford--Znajek
effect (BZ, Blandford \& Znajek 1977). If the angular momentum of
the central black hole is high enough, and a large scale poloidal field is generated and sustained
in the ergosphere, the BZ power could contribute substantially to the final jet/outflow power, but the
main conclusion of our investigation, namely the dependence of the jet power on the
disc accretion rate, would remain unchanged (see e. g. Meier 2001).} can be expressed as (Livio, Ogilvie \&
Pringle 1999)
\begin{equation}
\label{eq_lj}
L_{\rm j}=\left(\frac{B_{\rm p}^2}{8\pi}\right)2\pi R_{\rm cor}^2 R_{\rm cor}\Omega.
\end{equation}
Here $R_{\rm cor}$ is the size of the region where most of the coronal power
is dissipated and $\Omega$ is the typical angular velocity of the magnetic
field lines.
Maximal rotation of the central black hole
 may be important in this context more for its deeper potential well and the larger
rotational velocities that the accreting gas can attain (see also Meier 2001) than for the
extraction of rotational energy from the black hole itself.

The magnetic field that emerges from the disc into the corona must be mainly
azimuthal, as simulations also suggest \cite{ms00,mach00}.
If the disc is not threaded by any external large scale poloidal field, as it
is likely to be in the case of central black holes, the strength of the poloidal
field component depends on the typical scaleheight of a coronal magnetic flux tube
and on the capability of reconnection events to create larger
and larger coherent structures. Tout \& Pringle (1996) have proposed numerical and
analytic models in which stochastic reconnection creates magnetic loops on all scales,
with a power-law distribution.
Here, as the jet-launching region is identified with the corona itself, we consider
the constraints put on the poloidal field by the coronal geometry rather
than the disc one.

A magnetic flux tube emerging from the disc should have
a cross section of the order of the disc thickness \cite{grv79}. Nonetheless, the strong
differential rotation at the tube footpoints causes a twisting of the tube magnetic field
and increase its tension, counterbalancing its sideways expansion (see Parker (1979), \S 9.1, 9.6).
As both theoretical arguments and numerical simulations
have shown \cite{aly90a,aly90b,alat96,rom98},
the net effect of this energy
injection in the magnetic coronal loops is therefore
a rapid loop expansion, followed by reconnection
(at a height $H_a$ much larger than the disc thickness) and opening of the field lines.
Then, if $H_a$ is the typical coronal flux tube scaleheight (height of a reconnection site), we have
\begin{equation}
\label{eq_bp_b}
\frac{B_{\rm p}}{B} \simeq \frac{H_a}{R_{\rm cor}}.
\end{equation}
From eqs. (\ref{eq_b_lh}), (\ref{eq_lj}) and (\ref{eq_bp_b}) we obtain
\begin{equation}
\label{eq_lj_lh}
L_{\rm j}=\frac{3}{2}L_{\rm H} \left(\frac{c}{v_{\rm diss}}\right)
\left(\frac{H_a}{R_{\rm cor}}\right)^2\left(\frac{R_{\rm cor} \Omega}{c}\right),
\end{equation}
which in turn gives, for the fraction of coronal power that goes into the MHD jet/outflow
\begin{equation}
\label{eq_eta}
\eta_{\rm MHD}=\left(1+\frac{2}{3}\left(\frac{v_{\rm diss}}{c}\right)
\left(\frac{R_{\rm cor}}{H_a}\right)^2\left(\frac{c}{R_{\rm cor} \Omega}\right)\right)^{-1}.
\end{equation}

Therefore, from eq.~(\ref{eq_eta}), we can conclude that the jet/outflow power is stronger if:
\begin{itemize}
\item[(a)]{The dissipation speed is low. In this case magnetic field does not dissipate
effectively via reconnection and can reach high intensities, favouring
MHD outflow launching;}
\item[(b)]{The coronal scaleheight is large  with respect to the distance from the central source. This would help in increasing
the relative strength of the poloidal component of the magnetic field, that is 
the one ultimately responsible for the powering of the jet;}
\item[(c)]{The rotational velocity of the field lines is high. This is related to
the gravitational potential of the central black hole (maximally rotating
Kerr holes can increase the angular velocity of field lines relative to a distant 
observer due to the strong frame dragging effect, see e.g Meier 2001), 
and to the typical radius of
maximal coronal dissipation.}
\end{itemize}
As an example, for $v_{\rm diss} \simeq 0.01 c$, $R_{\rm cor}\sim 7 R_{\rm S}$, 
$H_a\sim 2R_{\rm S}$ and
$\Omega=\Omega_{\rm K}(R_{\rm cor})$, we obtain $\eta_{\rm MHD} \simeq 0.55$: the MHD jet can carry away a substantial fraction of the coronal (and accretion) power. 
The fraction of jet power 
$\eta_{\rm MHD}$ strongly depends on the typical height of the coronal active regions.
Interestingly, for a given active region size, $R_a$, the greater its distance from the disc,
 $H_a$, the more important
will be synchrotron radiation as a source of soft photons for Comptonization 
(Di Matteo, Celotti \& Fabian 1999; Merloni \& Fabian 2001b) as opposed to external radiation
from the disc. This immediately suggests that
sources with observed variable high frequency synchrotron radiation 
are strong candidates for harbouring powerful MHD driven jets.

\subsection{The jet power II. Thermally driven outflows}
\label{sec_th}

As already discussed at the beginning of section 3, a crucial uncertainty about
the energetics of powerful coronae is whether dissipation of magnetic energy 
preferentially heats the protons or the electrons. It seems very likely that 
the fraction of energy dissipated into the electrons is negligible at least for 
optically thin plasmas where magnetic pressure is smaller than gas pressure 
(Quataert \& Gruzinov  1999). In this case, as 
discussed in Di Matteo, Blackman \& Fabian (1997), 
a two-temperature equilibrium (with $T_{\rm p} \gg T_{\rm e}$) 
is easily established.

Let us therefore consider the case of a two-temperature corona (see also Janiuk \& Czerny 2000;
R\'o\.za\'nska \& Czerny 2000)
and assume that all the dissipated magnetic energy is dumped into
the protons
\begin{equation}
\label{eq_mag_heat}
\left(\frac{d\epsilon_{B}}{dt}\right)_{-}=\frac{B^2}{8\pi}\frac{v_{\rm diss}}{R_a}
=\left(\frac{d\epsilon_{i}}{dt}\right).
\end{equation}
The protons exchange energy with the electrons only via Coulomb collision, so that,
in a steady state,
\begin{equation}
\left(\frac{d\epsilon_{\rm i}}{dt}\right)=\left(\frac{d\epsilon_{\rm e}}{dt}\right)_{+},
\end{equation}
where $(\frac{d\epsilon_{\rm e}}{dt})_{+}$ is the rate of energy transfer from
ions to electrons due to Coulomb collisions between populations with
Maxwellian distributions, calculated using the Rutherford
scattering cross section (Stepney \& Guilbert 1983):
\begin{equation}
\left(\frac{d\epsilon_{\rm e}}{dt}\right)_{+}=\frac{3}{2}\frac{m_{\rm e}}{m_{\rm p}}
\frac{n_e n_i \sigma_{\rm T}(kT_{\rm i}-kT_{\rm e}) c}{K_2(1/\theta_{\rm e})K_2(1/\theta_{\rm i})} \ln\Lambda f(\theta_{\rm e},\theta_{\rm i}),
\end{equation}
where $T_{\rm e,i}$ are the electrons and ions temperatures, respectively,
$\theta_j=kT_j/m_j c^2$ ($j={\rm e, i}$) are dimensionless temperatures,
$\ln \Lambda \approx 20$ is the Coulomb logarithm, $K$s are the modified Bessel functions and
\[
f(\theta_{\rm e},\theta_{\rm i}) = \left[\frac{2(\theta_{\rm e}+\theta_{\rm i})^2+1}{\theta_{\rm e}+\theta_{\rm i}}K_1\left(\frac{\theta_{\rm e}+\theta_{\rm i}}{\theta_{\rm e} \theta_{\rm i}}\right)+ 2K_0\left(\frac{\theta_{\rm e}+\theta_{\rm i}}{\theta_{\rm e} \theta_{\rm i}}\right)\right].
\]

In order to calculate the electron temperature, we equate the electron heating rate
with the total cooling rate
$q_{\rm cool}=q_{\rm Comp}+q_{\rm synch}+q_{\rm CS}$, which includes inverse Compton
scattering of both disc photons and self-absorbed synchrotron emission produced in
the active region itself (see e.g. Di Matteo, Celotti \& Fabian 1997,
Wardzi\'nski \& Zdziarski 2000).

\begin{figure}
\psfig{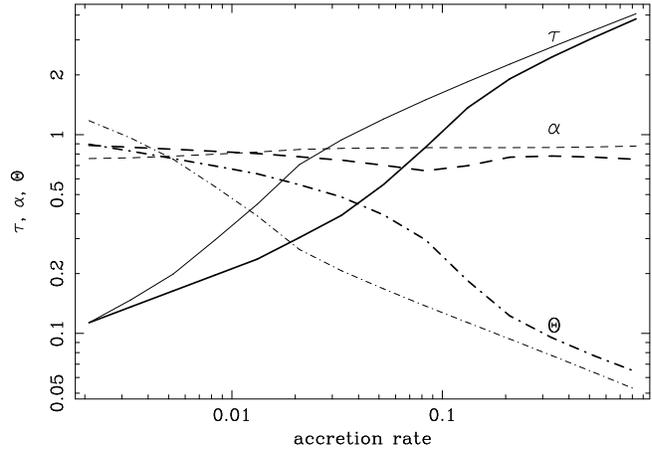}
\caption{Variation of the electron temperature in the corona
($\Theta=kT_{\rm e}/m_{\rm e}c^2$, dot-dashed lines) and of the X-ray spectral index
($\alpha$, dashed lines) as functions of the accretion rate, calculated for the
optical depth profile shown ($\tau$, solid lines). Thicker lines correspond to a central
 black hole mass $m=10$, thinner lines to $m=10^8$.
}
\label{tau_alpha_th}
\end{figure}

Once the heating rate is specified via eqs.~(\ref{eq_b_lh}) and (\ref{eq_mag_heat}), and the geometry of the corona is fixed, we can calculate electron and ion temperatures if we know the
coronal optical depth. Observationally, both black hole candidates in their Low/Hard state
and AGN have similar values of $\tau$, lying in a narrow range around
$\tau\sim 1$ \cite{gier97,zdz99}.
As an illustrative example, we made here the assumption that,
for each value of the central source mass, $M$,
the coronal optical depth is proportional to the accretion disc surface temperature.
In particular, for gas pressure dominated solution appropriate for low accretion rates,
we have chosen $\tau \simeq 4 \dot m^{2/5} (1-f)^{1/5}$. Therefore $\tau$ is
smaller for smaller accretion rates ($\tau \sim 0.2$ for $\dot m = 0.005$) and
we have $\tau \sim 1$ for $\dot m = 0.05 - 0.08$, depending on the black hole mass.
The dependence of $\tau$ on the accretion rate for central black holes of masses
$m=10$ and $m=10^8$ is shown in Fig.~\ref{tau_alpha_th}. It is important to
stress that the specific dependence of $\tau$ on the disc parameters we have assumed
here has been chosen only for illustrative purposes: the results we present on the
strength of thermally driven outflows would be qualitatively the same for any
physical situation in which colder and denser accretion discs were sandwiched by
lower optical depth coronae (as would be the case, for example, if the coronal optical depth
depended on the accretion disc ionization parameter $\xi=L_{\rm H}/nH_a^2$). 
In fact, the only requirement for the jet/outflow to dominate the energy 
budget at increasingly lower accretion rates is that the coronal 
optical depth decreases with decreasing accretion rate, regardless of the specific profile.
If instead $\tau$ was approximately independent of $\dot m$, the highest proton temperature
would be reached for $\dot m_{\rm cr}$, where the absolute coronal power, 
$L_{\rm c}=f\dot m L_{\rm Edd}$ is maximum. 

Once a value for
the coronal optical depth is fixed, for each value of the accretion rate we can then calculate
magnetic field intensity, electron and proton temperatures and the overall emitted
spectrum by solving the equation for the cooling and heating rate,
adopting an analytic approximation for the emissivities described in Merloni \& Fabian (2001b; Appendix A).
In Figure~\ref{tau_alpha_th} we also show, as functions of the accretion rate,
the calculated values of the
electron temperature $\Theta=kT_{\rm e}/m_{\rm e}c^2$ and spectral index $\alpha$
of the hard X-rays produced by inverse Compton scattering on the hot coronal electrons.

We stress that the values of the electron temperature depends also on the details
of the coronal geometry, namely, on the size, height and number of magnetic regions
reconnecting in the corona at any time \cite{ste95,dcf99}.
This is an inherently time dependent problem
(see e.g. Poutanen \& Fabian 1999, Merloni \& Fabian 2001b) and here we have chosen
average values for the coronal geometric parameters ($H_a=2 R_a=2 R_{\rm S}$)
that ensure that the
value of the spectral index is close to the observed one, and we are not interested in detailed
spectral modeling of the X-ray emission.

\begin{figure}
\psfig{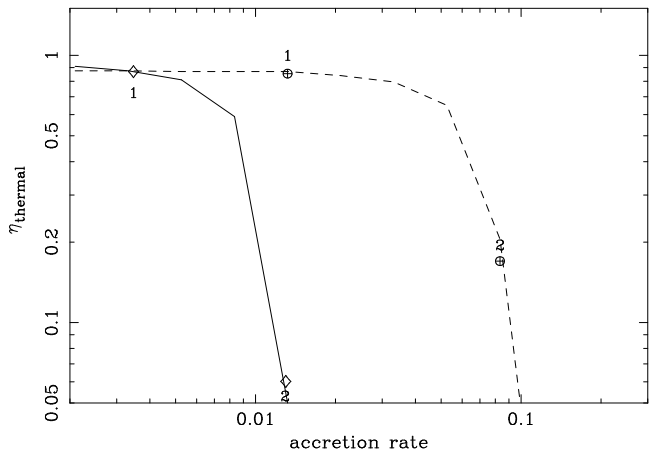}
\caption{The relative strength of thermally driven coronal outflow
as a function of the accretion rate for two different values of the central black hole
mass: $M=10^8 M_{\odot}$ (solid line) and $M=10 M_{\odot}$ (dashed line).
We have imposed maximal coronal strength by taking $K=0.99$. At the lowest accretion rates, where
$f\rightarrow K^{1/4}$ we have $\eta_{\rm thermal}\simeq 0.98$.}
\label{eta_mdot}
\end{figure}

If the coronal density is low enough, and if the power dissipated in the corona high
(and both conditions are more likely in low accretion rate discs, as we have shown
in section~\ref{sec_f_mdot}), the protons can be heated
up to temperatures higher than the virial one $T_{\rm vir}=GMm_{\rm p}/3kR_*$ (here
we may take as $R_*$ the radius at which the coronal dissipation is higher).
In this case the corona is likely to drive powerful jets/outflows \cite{pir77,lmt01}.
A fraction $\eta$ of the coronal power is advected away in the outflow in the form of bulk kinetic energy
and the heating rate of the protons is therefore reduced.
In such cases we may estimate a value for the parameter $\eta$ simply by
looking for
the minimal value of $\eta$ for which the protons are at the virial temperature.
The resulting values for such
$\eta_{\rm thermal}$ as a function of the accretion rate
for $K=0.99$ and two different black
hole masses are shown in Figure~\ref{eta_mdot}. The outflow/jet power
(kinetic energy plus Poynting flux) decreases sharply as the accretion rate increases and is
negligible as soon as the temperature up to which the protons are heated in the corona
drops below the virial one.

\section{Predicted spectra}
Regardless of the nature of the launching mechanism, a robust conclusion from the 
analysis of the previous sections is that when the fraction of gravitational power released
into a structured magnetic corona approaches unity, strong outflows ought to be produced.
Here we describe the basic spectral properties of the disc--corona emission. As a reference,
we will consider a sequence of models taken from the two--temperature solutions illustrated 
in Fig.~\ref{f_mdot_coeff} and \ref{eta_mdot}, but we emphasize that any of such  spectra would be observed 
if the jet was MHD driven, as far as it had the same value of $\eta$.   

For small and intermediate
values of $f$ (therefore for $\dot m \ga 0.08$ or $\dot m \ga 0.008$ for $m=10$ and
$m=10^8$, respectively) the seed soft photons for Comptonization come from the quasi-thermal
accretion disc emission (either intrinsic or reprocessed).
Then, the coronal cooling decreases when the accretion
rate decreases, and the X-ray spectrum gets harder, as can be seen if Fig.~\ref{tau_alpha_th}.
The hardest spectra are obtained for $f\rightarrow 1$ and $\eta < 0.8$. At this point
inverse Compton scattering on synchrotron photons produced in the corona
dominates the cooling.
Finally, as the ions become supervirial, more and more power is
carried away in the form of a powerful jet/outflow and the coronal heating is
reduced more rapidly than the cooling due to synchrotron emission and the spectrum gets
slightly softer again.

\begin{figure}
\psfig{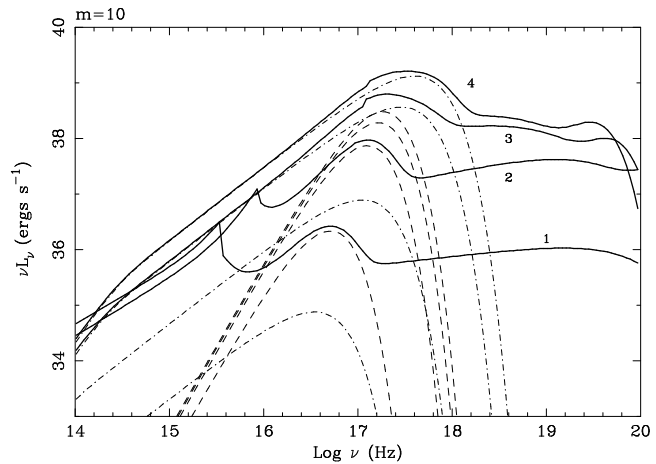}
\caption{The spectral energy distributions calculated for our thermally driven jet model
for $m=10$ (see text for details).
The four spectra correspond to the four points marked with
open circles in Fig.~\ref{f_mdot_coeff}. Solid lines show the total emerging spectra, while
dashed lines show the contribution of thermal reprocessed radiation from the disc
and dot-dashed lines the intrinsic multicolor disc emission (with outer radius
$R_{\rm out}= 2\times 10^5 R_{\rm S}$. The strong peak at
$10^{15}-10^{16}$ Hz represent the self absorbed synchrotron emission from the active
coronal regions: for spectrum 1  ($\dot m \simeq 0.01$) it clearly
represents the main source of seed photons for Comptonization. The 
total jet luminosity has been estimated assuming a flat spectrum 
($L_{\nu} \propto \nu^{\gamma}$ with $\gamma\sim 0$) extending up to the frequency
of coronal synchrotron self-absorption and a 10 per cent radiative
efficiency. No optically thin emission from the jet is included. Also neglected
is the reduction of the thermal reprocessed hump (dashed lines) that can be expected if
the X-ray emitting coronal active regions also move with the outflow at relativistic speeds
(Beloborodov 1999). This is by far the dominant contribution to the EUV/soft 
X-ray emission at low accretion rates.}
\label{spectra_10}
\end{figure}

\begin{figure}
\psfig{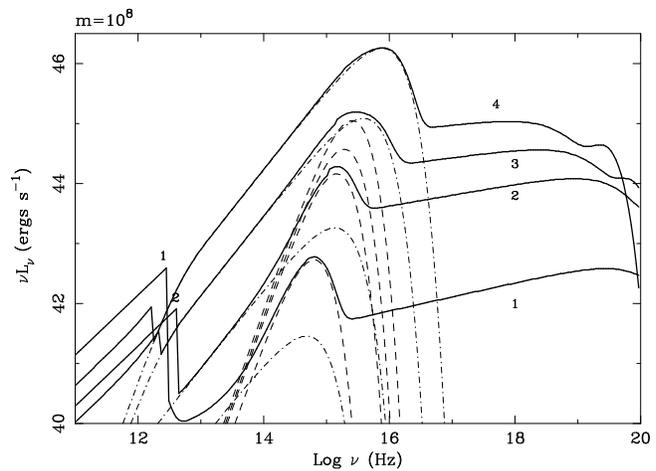}
\caption{The spectral energy distributions calculated for our thermally driven jet model
for $m=10^8$ (see text for details). Symbols and line styles are the same as in
Figure~\ref{spectra_10}. Here direct synchrotron radiation is always dominated by
the quasi-thermal disc component, due to the decreased value of the coronal magnetic
field ($B \propto m^{-1/2}$), except for spectrum 1, at the lowest accretion rate, where a strong 
jet component is evident up to about 1000 GHz. 
At accretion rates lower than what shown by curve 1 ($\dot m < 0.003$)
the shape of the SED does not change any longer, the overall luminosity being simply reduced.
}
\label{spectra_108}
\end{figure}

This is illustrated in Fig.~\ref{spectra_10} and \ref{spectra_108} which show
a sequence of spectral energy distributions for the stellar mass and supermassive black
hole cases, respectively.

We do not attempt to model in detail the emission from the jet/outflow. Given the high brightness temperatures
measured in the cores of some LLAGN, we can expect self-synchrotron Compton emission to be important
at low $\dot m$. Also, the high energy spectrum would be modified by even a small fraction
of non-thermal electrons in the corona and/or in the outflow (see e. g. Poutanen \& Coppi 1998;
Wardzi\'nski \& Zdziarski 2001).

\section{Discussion}
At the low accretion rates we are considering, both our corona dominated solution 
for a standard thin disc and ADAF/CDAF solution are admitted. 
A `strong ADAF principle' has been
advocated (Narayan \& Yi 1995) to ensure that the real systems will tend to
chose the ADAF/CDAF branch whenever possible. Here we have explored the possibility
that such a strong ADAF principle does not hold, and studied the expected properties (spectral
and energetic) of the alternative accretion mode. Nonetheless, there are admittedly some
similarities between the powerful outflow-dominated, radiatively-inefficient coronae
that we found here at the lowest accretion rates and a classical ADAF, and we might well be exploring
a region of the accretion flow parameter space where the two accretion modes are bridged together
(in this case mainly by virtue of strong magnetic field generation, rather than evaporation;
Meyer \& Meyer-Hofmeister 1994). Indeed, predictions from ADAF models
of a transition in spectral properties at $\dot m \la 0.1$ \cite{es97} are similar to those of our 
coronal outflow dominated sources, at least at high energies. 
Also, from our study we expect strong coronal outflows
at high values of the viscosity, similar to what is predicted by the 
theory of advective flows \cite{ia00}. 
The main testable difference between the two
models is the presence of a geometrically thick, cold, accretion disc in the inner part
of the flow. In the magnetic corona scenario, even at the lowest accretion rates, when
most of the accreting power is dissipated in the corona, the presence of the disc in the inner
part of the flow would manifest itself by reprocessing coronal radiation and producing
both a cold, quasi-thermal and a reflection components. Nonetheless, it is well known that
if the coronal luminosity is so high that the inner part of the accretion disc is
fully ionized, the reflection features are so weak that
it is difficult to distinguish between a truncated disc and an ionized one
\cite{dn01,brf01}. Furthermore, when the corona is so powerful that it produces relativistic outflows,
the bulk motion of the coronal outflowing material reduces the amount of high energy radiation
impinging on the disc, thus reducing the strength of the reprocessed components
\cite{bel99,mal00}.

When $f\rightarrow 1$ and $\eta\rightarrow 1$, the radiative efficiency of a coronal outflow
dominated flow depends on the radiative efficiency of the outflow itself. However,
unless $1-\eta \sim 10^{-3}$, we cannot achieve the extremely low efficiencies
($\epsilon \la 10^{-5}$) required to explain some nearby LLAGN \cite{dim01}. Such estimates, though, are based on the Bondi accretion rates estimated at relatively large distances from the source.
It is plausible to conceive that the strong outflows we envisage, by interacting
with the intergalactic medium, would stifle
accretion through the inner part of the disc
(an effect analogous to that of the ADIOS solution of Blandford \& Begelman 1999,
in the advective case), reducing the value of the accretion rate through the inner disc that we are
considering here and alleviating the need for too low a radiative efficiency
(see also discussions in Melia and Falcke, 2001, with regards to the Galactic center, 
and Di Matteo et al., 2001, for NGC 6166).

Our coronal-outflow-dominated solutions are both thermally and viscously
stable, as in general are all standard
Shakura--Sunyaev accretion disc solution in the gas pressure dominated regime.
Rapid and dramatic variability in the observed high-energy flux is nevertheless expected,
as X-rays are produced by  coronal structures that are the eventual outcome of
the turbulent magnetic field generation inside the disc. The geometry of these structures
(open vs. closed field lines, for example) plays a very important role and may
be such that, at times, parts of the corona become temporarily radiatively efficient.
The hard X-ray flaring event recently detected by Chandra from Sgr A$^*$ \cite{bag01}
might be an example of such an occasional coronal re-brightening. At the very low accretion
rates inferred for this source,
we predict the X-rays to be produced by inverse Compton scattering on synchrotron photons 
(see section 4), as indeed is suggested from the observations \cite{bag01}.
We note however that the interpretation of the galactic centre as a coronal-outflow-dominated
  disc is not straightforward. Firstly, the extreme faintness of the source
requires an accretion rate in the inner part of the flow extremely low compared to that
expected from simulations of the multiple wind sources in the inner $\sim 0.06$ pc 
\cite{cm97} and values of 
$\eta$ very close to unity. The former requirement of small $\dot m$ 
(also consistent with the polarization characteristics of SGR A$^*$ at 
millimeter and submillimeter wavelength, see  Agol, 2000, Quataert \& Gruzinov, 2000b, and 
Melia \& Falcke, 2001, and references therein)
may be due to the feedback from the coronal outflow 
suppressing the accretion rate at larger radii.
However, a second problem faced by our model is the over-prediction of the IR emission
that should be produced by the infalling wind impacting on the outermost part of the cold disc
\cite{fm97,cok99}. Apart from the possibility that the above mentioned interaction between
 outflowing and inflowing gas modify substantially the kinematics of the accretion flow, 
it is possible that in fact the inflowing matter 
forms a Keplerian disc only relatively close to
the central black hole \cite{mlc01}. Such a small disc would still be dominated 
by a strong outflowing corona as the one we have presented here, with enough power to
produce a jet responsible for the radio emission, as in the models of 
Falcke \& Markoff (2000). 
Clearly, the challenge posed by the
wealth of data from the galactic centre to any accretion model requires  
more detailed work than the one we have outlined here, which would be beyond the 
aim of this paper.

An important related issue concerns the fate of the cold thin disc when we extrapolate
our solution to extremely low accretion rates.
As we have discussed already in section \ref{sec_f_mdot}, very strong outflowing coronae might
have relevant effects on the nature of the angular momentum transport in the 
flow, and therefore on the disc structure.
The constant improvement of numerical simulations in recent years, make us believe that
the relevance of the solutions we describe here will be assessed in the near future.
At present, numerical simulations of MRI unstable accretion discs with sufficient
extension in the vertical direction seem able to generate only modestly powerful coronae
($f\sim$ 20 per cent, Miller \& Stone 2000). The inclusion of radiation in the MHD codes
\cite{ts01}, and a detailed study of the dependence of the solution on the accretion rate
 will prove crucial to settle the issue. 

We have discussed separately two different outflow--producing  mechanisms (MHD and thermal). 
Their relative importance crucially depends on the uncertain nature of the magnetic 
energy dissipation in the corona and of particle heating. However, as we have discussed in section 3, according to our current theoretical understanding of these issues, at least one 
of the two mechanisms must be at work in the physical conditions we envisage for a 
powerful corona. 
Clearly both may be at work
at the same time, and influence each other. For example, when the conditions for MHD jet launching
are favourable, and a larger number of open field lines thread the inner part of the corona,
then, depending on the inclination angle of the field lines, a larger mass can be
channeled into the outflow, reducing the coronal density and therefore the Coulomb cooling
of the hot ions and enhancing the thermally driven jet power.

Radiatively inefficient flows are also expected at high accretion rate,
when the flow is geometrically and optically thick and
the radiation produced is trapped by the flow and advected with it \cite{kat77,bm82,bel98}.
When the accretion rate is of the order of the Eddington one,
the disc thickness stays moderate and a vertically integrated  approximation may be retained
(slim discs, Abramowicz et al. 1988).
In the limit $\dot m \gg 1$, though,
the behaviour of the disc still remains an open issue. The problem is inherently 2D \cite{ia00}, and
the simultaneous roles of convection, advection and outflows have to be assessed in order to
properly model the expected SED. If a strong outflow was produced, through a magnetic corona
analogous to the one we envisaged for low accretion rates, or through any other radiative or
hydrodynamical
mechanism, the total amount of outflowing kinetic power could exceed the Eddington limit (and be
much larger than the radiated one; Meier 1982).
We just remark here that observationally, black holes
believed to accrete at super-Eddington rates bear many similarities with their low-accretion rate
counterparts. This class of object may include GBHC in the very high state
(see Done 2001 and references therein), whose X-ray spectra
show a power-law component usually as strong as the quasi-thermal one;
the microquasar GRS 1915+105 \cite{bmf00}; broad line radio galaxies
(BLRG, see e. g. Zdziarski 1999 and reference therein);
powerful blazars (see in particular Ghisellini \& Celotti 2001a, where it is shown how
the relativistic jet in these sources may dominate the total output power);
and possibly, narrow line Seyfert 1s (NLS1,
Mineshige et al. 2000). Most likely, newly formed supermassive black holes at the epoch of
galaxy formation and quasars in their early evolutionary stages might be fed at
super-Eddington rates, therefore a better understanding of accretion flows and jet/outflow
production in these sources will shed light on the processes of galaxy formation and evolution
\cite{fab99}.

\section{Conclusions}
We studied the energetics and power output of magnetic structured coronae generated by
geometrically thin, optically thick accretion discs at low accretion rates.
We have assumed that angular momentum transport in the disc is due to magnetic
turbulent stresses. The magnetic field, amplified by MRI, saturates because of the strong buoyancy of flux tubes.
Then, we have shown that both magnetic energy density and effective viscous stresses inside the disc
are proportional to the geometric mean of the total (gas plus radiation) and gas pressure. 
In this case, the relative strength
of the corona increases as the accretion rate decreases.
We have then studied the energetics of the corona itself, and assessed the relevance of the various mechanisms
that can generate powerful outflows from the coronal region.
We have shown that, depending on the actual coronal field geometry, MHD launching mechanisms can indeed
produce strong outflows, whose power is of the order of the radiated one. Moreover, if the corona
is two-temperature, a thermally driven outflow can easily dominate the power output from the source
at low accretion rates.
Thus, if the jet/outflow do not radiate efficiently, the whole system
can be very under-luminous compared to a standard radiatively efficient accretion disc.
The model we have presented might be relevant to the accretion mode of
 low-luminosity black holes, either GBHC in the {\it low/hard} state or LLAGN observed in the nuclei of 
nearby galaxies.

\section*{Acknowledgments}
The manuscript has been improved due to the comments and criticisms  
 of the referee, Prof. Ramesh Narayan, and to the many suggestions 
from Annalisa Celotti, whom we both thank.
This work was done in the research network
``Accretion onto black holes, compact stars and protostars"
funded by the European Commission under contract number
ERBFMRX-CT98-0195'. AM and ACF thank the PPARC and the Royal Society
for support, respectively.

\bsp

\label{lastpage}

\end{document}